\documentclass[prb,superscriptaddress, 12pt]{revtex4}

\usepackage{bm,amsmath,amssymb,graphicx}

\begin{document}

\title{Gravitation from a quantum mechanical argument: phase space compression}

\author{Victor Atanasov}
\affiliation{Faculty of Physics, Sofia University, 5 blvd J. Bourchier, Sofia 1164, Bulgaria}
\email{vatanaso@phys.uni-sofia.bg}

\begin{abstract} 
We use a quantum mechanical charged particle as a test particle which probes the dynamics of force-related fields it is subject to. We allow for geodesic motion and relations involving gravitation appear. Gravitation affects quantum dynamics by modifying operator algebra. The emerging commutator between momentum's components is recognized as being proportional to electromagnetic field strength tensor. We define electromagnetic field sources through momentum's components commutator which is proportional to geometric (gravitational) quantities. As a result, a source of the electric field can be thought of as the geometric disturbance. The framework points to the non-existence of mass-less charges and gravitation being able to introduce compressability of the quantum mechanical system's phase space, which constitutes its main coupling to the quantum (condensed matter) system. 
\end{abstract}

\maketitle

\section{Introduction}

Feynman showed Dyson a derivation which started with
the cannonical commutation relations for a single non-relativistic
quantum particle obeying Newton's law of motion and deduced the existence of electro-magnetic field satisfying Maxwell's equations\cite{dyson}. However, since this proof lead to nothing, previously unknown, Feynman abandoned seeking its publication. Feynman's proof is self-consistent but lacks relativistic covariance\cite{covariance}. Later generalizations (to non-Abelian fields and Yang-Mills equations as well\cite{nonabelian}) overcame this weakness and now we believe the original argument is complete\cite{bib, bib2}.

In this paper we attempt to ressurect Feynman's original argument (a ''joke'' in his own words\cite{joke}) with the sole intention of exploring its profound physical insight: {\it one can (partially) extract the equations of motion for the force-fields a quantum particle is subject to using the quantum particle as a test particle to these fields.} Note, electromagnetism was entirely constructed with the help of the notion of a ''test charge'' (one may call that approach ''empirical''). On the other hand, the theory of gravitation is not constructed in an empirical manner. It is constructed via thought experiments and with the help of the equivalence principle which leads to all particles (including the ''test'' ones) freely falling, namely following geodesics and being totally oblivious of the gravitational interaction (pure curvature of the embedding space-time). General theory of relativity postulates the matter-field interactions, rather than merely describing the interactions with the curvature field, test particles are probing.  

We believe, Feynman's original argument contains the possibility of recognizing a quantum particle (system) as actually being a test particle that can probe the dynamics of the underlying force-fields and as a result, if applied to the gravitational field, to act as a legitimate ''test particle'' revealing gravitational field's dynamics and field - quantum matter interaction outside the macro-scopic, self-consistent but postulative General theory of relativity\cite{einstein}. 

The paper is organized as follows: section \ref{sec.motivation} sets the matematical form of the idea that quantum particle is able to act as a test particle probing the dynamics of the underlying force-fields. Section  \ref{sec.equations} contains the set of matematical equations force-fields are subject to. Here we also explicitly write the non-vanishing momentum components' commutator. Next section \ref{sec.Analysis} identifies the force-fields for what they are: electric, magnetic and gravitational (affine connection components). Section \ref{sec.consequences} recognizes the effect of the geometric field on the phase space occupied by an eigenstate. Next section \ref{sec.commutator} discusses the physical meaning of $\left[ {p}^{\mu}  ,   {p}^{\nu}  \right],$ which turns out to be proportional to the electromagnetic field strength tensor. We end the paper with section \ref{sec.G-QM int} which discusses the quantum matter's interaction with the curved geometry. 

\section{Motivation}\label{sec.motivation}

Suppose the newtonian force experienced by a quantum particle is given by
\begin{eqnarray}\label{F}
f^{j}=m \ddot{x}^{j} = {f_0}^{j} + {f_1}^{j}_{\mu}\dot{x}^{\mu} + {f_2}^{j}_{\mu \nu}\dot{x}^{\mu} \dot{x}^{\nu}
\end{eqnarray}
where the individual components ${f_0}^{j}$, ${f_1}^{j}_{\mu}$ and ${f_2}^{j}_{\mu \nu}$ are force-related fields which dynamics is unknown at present. Next, suppose we impose the following commutation relations which are expressions of localizability:
\begin{eqnarray}\label{[x,x]}
\left[  x^{\mu} ,  x^{\nu}  \right]=0
\end{eqnarray}
and quantum mechanical uncertainty
\begin{eqnarray}\label{[x, dot x]}
m \left[  x^{\mu} ,  \dot{x}^{\nu}  \right]=i \hbar g^{\mu \nu}\approx i \hbar  \delta^{\mu \nu} + i \hbar  h^{\mu \nu} + o(h^2)
\end{eqnarray}
where $g^{\mu \nu}$ is the induced metric of the three-dimensional section of the space-time onto which the quantum particle (system) abides. Note, the r.h.s. of the latter commutator differs from the standard Kronecker delta $\delta^{\mu \nu}$ of the canonical commutation relations due to our desire to allow for extra degrees of freedom associated with the gravitational filed. These field degrees of freedom should find their way into the definition/dynamics of the force-related fields in (\ref{F}). In the weak curvature limit $| h^{ij} | \ll 1$ relation (\ref{[x, dot x]}) coincides with the standard quantum mechanical commutator. This form of the commutator has been explored with the conclusion that the only possible fields that can consistently act on a quantum mechanical particle (system) are scalar, gauge and gravitational fields\cite{Tanimura}.

The quintessence of what is to follow is an attempt to extract the governing dynamical equations the force-related fields ${f_0}^{j}$, ${f_1}^{j}_{\mu}$ and ${f_2}^{j}_{\mu \nu}$ are subject to. In a sense, we will use the quantum mechanical system as a test particle which behaviour should contain the governing equation (or at least part of them) that relate ${f_0}^{j}$, ${f_1}^{j}_{\mu}$ and ${f_2}^{j}_{\mu \nu}$. Thus, previously unnoticed relations involving gravitation may appear.

Indeed, assuming a newtonian force which acts on a quantum mechanical system is debatable. However, in the case of a quantum condensate, such an assumption is not in disagreement with the empirical behaviour. The issue with the validity of the results to follow in the highly curved space-time regime can be settled with a reference to the equivalence principle, namely the geodesic motion is a property of gravitational interaction which does not stem from the particular form of the equations governing the fields' dynamics.

\section{Governing Equations} \label{sec.equations}

First, let us build a vocabulary of relations to further use in calculations: 1.) let us apply total time derivative to (\ref{[x,x]}), then
\begin{eqnarray}\label{symmetry from [x,x]}
\left[  x^{\mu} ,  \dot{x}^{\nu}  \right]= \left[  x^{\nu} ,  \dot{x}^{\mu}  \right]
\end{eqnarray}
holds; 2.) let us apply total time derivative to (\ref{symmetry from [x,x]}), then $
\left[  \dot{x}^{\mu} ,  \dot{x}^{\nu}  \right]+ \left[  x^{\mu} ,  \ddot{x}^{\nu}  \right]=
\left[  \dot{x}^{\nu} ,  \dot{x}^{\mu}  \right]+ \left[  x^{\nu} ,  \ddot{x}^{\mu}  \right]
$ holds. In other words, the combination 
$\left[  \dot{x}^{\mu} ,  \dot{x}^{\nu}  \right]+ \left[  x^{\mu} ,  \ddot{x}^{\nu}  \right]$
is symmetric with respect to index exchange; 3.) The total time derivative of (\ref{[x, dot x]}) yields
\begin{eqnarray}\label{first [dot x, dot x]}
m \left[  \dot{x}^{\mu} ,  \dot{x}^{\nu}  \right] +  \left[  x^{\mu} , m \ddot{x}^{\nu} \right] =i \hbar \left( \partial_{t} g^{\mu \nu} + \partial_{j} g^{\mu \nu} \, \dot{x}^{j}    \right).
\end{eqnarray}
Given the above point 2, both l.h.s. and r.h.s. are symmetric with respect to index 
exchange. On one hand, this last relation provides a mean to express the momentum commutator $ \left[  p^{\mu} ,  p^{\nu}  \right]/m = m \left[  \dot{x}^{\mu} ,  \dot{x}^{\nu}  \right]$ in terms of the force fields and the metric. In effect, using (\ref{F}), (\ref{[x,x]}) and (\ref{[x, dot x]}) yields
\begin{eqnarray}\label{first [dot x, dot x]}
\nonumber m \left[  \dot{x}^{\mu} ,  \dot{x}^{\nu}  \right] &=&  i \hbar \partial_{t} g^{\mu \nu} -  i \hbar m^{-1} {f_1}^{\nu}_{j}  g^{\mu j} - m^{-2}  \hbar^2 {f_2}^{\nu}_{jk} \partial^{j}  g^{\mu k}\\
&&- i \hbar m^{-1} {f_2}^{\nu}_{jk}\left( g^{\mu j}\dot{x}^{k}  + g^{\mu k}\dot{x}^{j} 
\right)+i\hbar \partial_{j} g^{\mu \nu} \, \dot{x}^{j}
\end{eqnarray}

On the other hand, the Jacobi identity (a virtue of the commutator) leads to yet another expression. Indeed, $
 \left[ {x}^{\rho}    ,  \left[  \dot{x}^{\mu}  ,  \dot{x}^{\nu}  \right] \right]
+  \left[ \dot{x}^{\nu}    ,  \left[ {x}^{\rho} ,  \dot{x}^{\mu}  \right] \right] 
+ \left[ \dot{x}^{\mu}     ,  \left[ \dot{x}^{\nu}, {x}^{\rho}  \right] \right] =0,  
$ therefore
\begin{eqnarray}\label{Jacobi1}
\left[ {x}^{\rho}    , m  \left[  \dot{x}^{\mu}  ,  \dot{x}^{\nu}  \right] \right] &=&  m^{-1} (i \hbar)^2 \left(  \partial^{\nu}  g^{\rho \mu}  
- \partial^{\mu} g^{\rho \nu}  \right).
\end{eqnarray}
The r.h.s. of the last relation is only a function of the coordinates, which is an indication of the most general form $m  \left[  \dot{x}^{\mu}  ,  \dot{x}^{\nu}  \right]$ can take. It is
\begin{eqnarray}\label{form [dot x, dot x]}
m  \left[  \dot{x}^{\mu}  ,  \dot{x}^{\nu}  \right] &=& 
F^{\mu \nu} + G^{\mu \nu \rho} \dot{x}_{\rho}.
\end{eqnarray}
Here $F^{\mu \nu}$ and $G^{\mu \nu \rho}$ are functions of the coordinates and time only. Their dimensionality is $
\left[ F^{\mu \nu} \right] = J $ and $ \left[ G^{\mu \,  \nu \rho} \right] =  {J}.{(m/s)}^{-1}.
$ Furthermore, $
 \left[ {x}^{\rho}    , m  \left[  \dot{x}^{\mu}  ,  \dot{x}^{\nu}  \right] \right] =  m^{-1} i \hbar G^{\mu \, \nu \rho}
$
which means that
\begin{eqnarray}\label{G_mu_nu_rho}
G^{\mu \,  \nu \rho} &=& i \hbar \left(  \partial^{\nu}  g^{\rho \mu}  - \partial^{\mu} g^{\rho \nu}  \right)
\end{eqnarray}
and $G^{\mu \,  \nu \rho}= G^{\mu \,  \rho \nu }$, that is the set of symmetries completes with
$F^{\mu \nu}=-F^{\nu \mu },$ $
G^{\mu \,  \nu \rho}= - G^{\nu \, \mu \rho}$ and
$G^{\mu \,  \nu \rho}= G^{\mu \,  \rho \nu }.$

Combining the respective quantities from (\ref{first [dot x, dot x]}), (\ref{form [dot x, dot x]}) and (\ref{G_mu_nu_rho}) we obtain
\begin{eqnarray}\label{eq.f2}
 {f_2}^{\nu\, \mu \rho}
 &=&m \frac12 \left( \partial^{\rho} g^{\mu \nu} + \partial^{\mu}  g^{\rho \nu}  -\partial^{\nu} g^{\rho \mu}  \right)=m \Gamma^{\nu\, \mu \rho}\\\label{eq.f1}
  i \hbar m^{-1} {f_1}^{\mu \nu} &=& F^{\mu \nu} + \frac{\hbar^2}{2m} \left(  \Gamma^{\nu}_{\; jk} \partial^{j}  g^{\mu k} - \Gamma^{\mu}_{\; jk} \partial^{j}  g^{\nu k} \right) \quad {\rm anti-symmetric \; part}\\\label{eq.dt g}
 i \hbar \partial_{t} g^{\mu \nu} &=& \frac{\hbar^2}{2m} \left(  \Gamma^{\nu}_{\; jk} \partial^{j}  g^{\mu k} + \Gamma^{\mu}_{\; jk} \partial^{j}  g^{\nu k} \right) \quad {\rm symmetric \; part}
\end{eqnarray}
where $\Gamma^{\nu\, \mu \rho}$ is the Christoffel symbol of the first kind.

Taking total time derivative of (\ref{form [dot x, dot x]}) and 
multiplying with the Levi-Civita symbol we obtain:
\begin{eqnarray}\label{}
\nonumber 2 \epsilon_{\lambda \mu \nu} m  \left[  \dot{x}^{\mu}  ,  \ddot{x}^{\nu}  \right]  &=& \epsilon_{\lambda \mu \nu} \left( \partial_{t} F^{\mu \nu} + m^{-1} G^{\mu \nu \rho} {f_0}_{\rho} \right) + \\
\nonumber && \epsilon_{\lambda \mu \nu} \left( \partial^{\rho} F^{\mu \nu}  +  \partial_{t} G^{\mu \nu \rho} + m^{-1}  G^{\mu \nu \sigma} {f_1}^{\; \rho}_{\sigma}   \right) \dot{x}_{\rho}+\\
&& \epsilon_{\lambda \mu \nu} \left( \partial^{\sigma} G^{\mu \nu \rho} + m^{-1} G^{\mu \nu \tau} {f_2}_{\tau}^{\;  \sigma \rho}
\right) \dot{x}_{\sigma} \dot{x}_{\rho} 
\end{eqnarray}
On the other hand
\begin{eqnarray}\label{}
\nonumber 2 \epsilon_{\lambda \mu \nu} m  \left[  \dot{x}^{\mu}  ,  \ddot{x}^{\nu}  \right]  &=&  - 2 \epsilon_{\lambda \mu \nu} \frac{i\hbar}{m} \left( \partial^{\mu}{f_0}^{\nu} + \partial^{\mu} {f_1}^{\nu \sigma}\dot{x}_{\sigma} + \partial^{\mu} {f_2}^{\nu\, \rho \sigma}\dot{x}_{\rho} \dot{x}_{\sigma}    \right)\\
\nonumber && - 2 \epsilon_{\lambda \mu \nu} \left\{ m^{-1} {f_1}^{\nu \sigma} F_{\sigma}^{\; \mu} + m^{-1} {f_1}^{\nu \sigma} G_{\sigma}^{\; \mu \rho} \dot{x}_{\rho}+\right.\\
&&m^{-1} {f_2}^{\nu\, \rho \sigma} \left( F_{\rho}^{\; \mu} \dot{x}_{\sigma} + F_{\sigma}^{\; \mu}\dot{x}_{\rho}  - i \hbar m^{-1} \partial_{\rho} F_{\sigma}^{\; \mu}   \right)+\\
\nonumber &&\left. +m^{-1} {f_2}^{\nu\, \tau \rho} G_{\tau}^{\; \mu \sigma} \dot{x}_{\sigma}\dot{x}_{\rho} + m^{-1} {f_2}^{\nu\, \tau \sigma}  G_{\tau}^{\; \mu \rho} \dot{x}_{\sigma} \dot{x}_{\rho} - i \hbar m^{-1} {f_2}^{\nu\, \tau \sigma} \partial_{\sigma} G_{\tau}^{\; \mu \rho} \dot{x}_{\rho} \right\}
\end{eqnarray}
Equating the r.h.s. of the two equations leads to the following set of equations
\begin{eqnarray}\label{eq.dt F}
\nonumber \epsilon_{\lambda \mu \nu} \left( \partial_{t} F^{\mu \nu} + m^{-1} G^{\mu \nu \rho} {f_0}_{\rho} \right) &=& 2 \epsilon_{\lambda \mu \nu} \left\{  \frac{i \hbar} {m^{2}} {f_2}^{\nu\, \rho \sigma}   \partial_{\rho} F_{\sigma}^{\; \mu} \right.\\
&&\qquad \left.  - \frac{i\hbar}{m}  \partial^{\mu}{f_0}^{\nu} - \frac{ {f_1}^{\nu \sigma} F_{\sigma}^{\; \mu} }{m} \right\}   \\\label{eq.d_rho F}
\nonumber \epsilon_{\lambda \mu \nu} \left( \partial^{\rho} F^{\mu \nu}  +  \partial_{t} G^{\mu \nu \rho} + m^{-1}  G^{\mu \nu \sigma} {f_1}^{\; \rho}_{\sigma}   \right) &=& - 2  \epsilon_{\lambda \mu \nu} \left\{ \frac{i\hbar}{m}  \partial^{\mu} {f_1}^{\nu \rho} + m^{-1} {f_1}^{\nu \sigma} G_{\sigma}^{\; \mu \rho}    \right.\\
&& \quad  \left. + 2 m^{-1} {f_2}^{\nu\, \rho \sigma} F_{\sigma}^{\; \mu}   -  i\hbar m^{-2} {f_2}^{\nu\, \tau \sigma}  \partial_{\sigma} G_{\tau}^{\; \mu \rho}
 \right\}\\\label{eq.d_rho G}
\nonumber \epsilon_{\lambda \mu \nu} \left( \partial^{\sigma} G^{\mu \nu \rho} + m^{-1} G^{\mu \nu \tau} {f_2}_{\tau}^{\; \rho \sigma}
\right) &=& - 2  \epsilon_{\lambda \mu \nu}\left\{ \frac{i\hbar}{m}  \partial^{\mu} {f_2}^{\nu\, \rho \sigma}  +m^{-1} {f_2}^{\nu\, \tau \rho} G_{\tau}^{\; \mu \sigma} \dot{x}_{\sigma}\dot{x}_{\rho} \right.\\
&&\qquad \left.+ m^{-1} {f_2}^{\nu\, \tau \sigma}  G_{\tau}^{\; \mu \rho} \dot{x}_{\sigma} \dot{x}_{\rho}    \right\}
\end{eqnarray}
which are rather involved. Yet other dynamic relations come from the following Jacobi identity: $\epsilon_{\lambda \mu \nu}\left[  \dot{x}^{\lambda}  , m  \left[  \dot{x}^{\mu}  ,  \dot{x}^{\nu}  \right] \right]=0.$ They are
\begin{eqnarray}\label{eq. dirac}
i \hbar \epsilon_{\lambda \mu \nu} \partial^{\lambda} F^{\mu \nu} &=&
\epsilon_{\lambda \mu \nu} G^{\mu \nu \rho} g_{\rho \sigma} 
F^{\lambda \sigma} \\\label{eq. G=G}
i \hbar \epsilon_{\lambda \mu \nu} \partial^{\lambda} G^{\mu \nu}_{\;\;\; \sigma}&=&  - i \hbar \epsilon_{\lambda \mu \nu} G^{\mu \nu \rho} \partial^{\lambda} g_{\rho \sigma}  + 
\epsilon_{\lambda \mu \nu} G^{\mu \nu}_{\;\;\;\tau} G^{\lambda  \tau}_{\;\;\; \sigma}
\end{eqnarray}

\section{Analysis}\label{sec.Analysis}

To fully understand the meaning of the governing equations and the field dynamics they encode we will explore the vanishing gravitational field limit, that is $ h^{\mu \nu} \to 0.$ Then (\ref{F}) reduces to
\begin{eqnarray}\label{force_g=0}
f^{j}=m \ddot{x}^{j} = {f_0}^{j} + {f_1}^{j}_{\mu}\dot{x}^{\mu},
\end{eqnarray}
relation (\ref{[x,x]}) remains as it is, (\ref{[x, dot x]}) reduces to the standard canonical commutator $
m \left[  x^{\mu} ,  \dot{x}^{\nu}  \right]=i \hbar \delta^{\mu \nu}, $
the non-commutativity of momentum's components (\ref{form [dot x, dot x]}) reduces to
$
m  \left[  \dot{x}^{\mu}  ,  \dot{x}^{\nu}  \right] = 
F^{\mu \nu}.$
In effect, (\ref{eq.dt g}) and (\ref{eq.f2}) reduces to the trivial $0=0,$ while (\ref{eq.f1}) turns into
\begin{eqnarray}\label{eq.f1_g=0}
  i \hbar m^{-1} {f_1}^{\mu \nu} &=& F^{\mu \nu} .
\end{eqnarray}
Furthermore, (\ref{eq.d_rho F}),  (\ref{eq.d_rho G}) and (\ref{eq. G=G}) identically vanish, while (\ref{eq.dt F}) turns into
\begin{eqnarray}\label{eq.dt F_g=0}
\epsilon_{\lambda \mu \nu} \partial_{t} F^{\mu \nu}  &=& 2 \epsilon_{\lambda \mu \nu} \left\{   - \frac{i\hbar}{m}  \partial^{\mu}{f_0}^{\nu} - \frac{ {f_1}^{\nu \sigma} F_{\sigma}^{\; \mu} }{m} \right\}  
\end{eqnarray}
and (\ref{eq. dirac}) into
\begin{eqnarray}\label{eq. dirac_g=0}
i \hbar \epsilon_{\lambda \mu \nu} \partial^{\lambda} F^{\mu \nu} &=& 0
\end{eqnarray}
Relation (\ref{F_mu nu}) suggests that $F^{\mu \nu}$ is anti-symmetric and can take the form
\begin{eqnarray}\label{1 F_mu nu=}
F^{\mu \nu} &=& {\rm C} \left( \partial^{\mu} A^{\nu} - \partial^{\nu} A^{\mu} \right)
\end{eqnarray}
or
\begin{eqnarray}\label{2 F_mu nu=}
F^{\mu \nu} &=& {\rm C}  \epsilon^{\mu \nu \tau} B_{\tau}
\end{eqnarray}
Inserting (\ref{2 F_mu nu=}) into (\ref{eq. dirac_g=0}) yields the absence of magnetic charge condition
\begin{eqnarray}\label{}
 \partial^{\lambda}  B_{\lambda}={\rm div} \vec{B} = 0,
\end{eqnarray}
which has the standard solution $B_{\lambda}= \epsilon^{\lambda \mu \nu}\partial_{\mu}  A_{\nu}.$ Thus (\ref{1 F_mu nu=}) is confirmed and (\ref{eq.dt F_g=0}), by virtue of the vanishing of the last term due to symmetry, turns into 
\begin{eqnarray}\label{1 eq.dt F_g=0}
{\rm C} 2  \partial_{t} B_{\lambda}  &=& - 2 \frac{i\hbar}{m}  \epsilon_{\lambda \mu \nu}   \partial^{\mu}{f_0}^{\nu}  
\end{eqnarray}
which is identified as the Faraday's law of electromagnetic induction, therefore with the help of (\ref{force_g=0})
\begin{eqnarray}\label{f_0}
{f_0}^{\nu}=q E^{\nu},  
\end{eqnarray}
where $E^{\nu}$ are the components of the electric field, $q$ is the charge of the particle and 
\begin{eqnarray}\label{C}
{\rm C}   &=&  \frac{i\hbar }{m} q.   
\end{eqnarray}
Clearly ${f_0}^{\nu}$ in (\ref{force_g=0}) is the electric force acting on a charged particle in an external electric field. 
Now combining (\ref{eq.f1_g=0}) and (\ref{2 F_mu nu=})
\begin{eqnarray}\label{f1=Bx}
 {f_1}^{\mu \nu} &=&  q \epsilon^{\mu \nu \tau} B_{\tau}=q \left( \partial^{\mu} A^{\nu} - \partial^{\nu} A^{\mu} \right) .
\end{eqnarray}
reveals the true meaning of the second term in (\ref{force_g=0}) which is the magnetic force ${f_1}^{j}_{\mu}\dot{x}^{\mu}=q \left( \vec{B} \times \vec{\dot{x}}  \right)^j$  acting on a charged particle in an external magnetic field.  

Having cleared the meaning of the first two force terms in (\ref{F}) we are ready to reveal the last one. For the purpose, we have to abandon the vanishing gravitational field limit. In this case, combining (\ref{F}) with (\ref{eq.f2}), (\ref{f_0}) and (\ref{f1=Bx}) produces a force term of the form
\begin{eqnarray}\label{F_revealed}
f^{j}=m \ddot{x}^{j} = q E^{j} + q \epsilon^{j \nu \mu} B_{\nu} \dot{x}^{\mu} + m \Gamma^{j}_{\mu \nu}\dot{x}^{\mu} \dot{x}^{\nu}.
\end{eqnarray}
Here the last term can be identified as a ''geodesic force''. Note, this particular form of the last term was only possible due to the assumption that the canonical commutator is proportional to the metric.

Now the defining equations (\ref{eq.f1}), (\ref{eq.dt g}) and (\ref{form [dot x, dot x]}) become
\begin{eqnarray}\label{F_mu nu defined}
 F^{\mu \nu} &=& - \frac{ i \hbar}{ m}  \, q \left( \partial^{\mu} A^{\nu} - \partial^{\nu} A^{\mu} \right) - \frac{\hbar^2}{2m} \left(  \Gamma^{\nu}_{\; jk} \partial^{j}  g^{\mu k} - \Gamma^{\mu}_{\; jk} \partial^{j}  g^{\nu k} \right) \\
 i \hbar \partial_{t} g^{\mu \nu} &=& \frac{\hbar^2}{2m} \left(  \Gamma^{\nu}_{\; jk} \partial^{j}  g^{\mu k} + \Gamma^{\mu}_{\; jk} \partial^{j}  g^{\nu k} \right)\\
m  \left[  \dot{x}^{\mu}  ,  \dot{x}^{\nu}  \right] &=& 
F^{\mu \nu} + i \hbar \left(  \partial^{\nu}  g^{\rho \mu}  - \partial^{\mu} g^{\rho \nu}  \right) \dot{x}_{\rho}.
\end{eqnarray}

\section{Consequences from gravitational modification of operator algebra}\label{sec.consequences}

The equivalence principle expresses both the equality of gravitational and inertial mass, and the Einstein's revelation that the gravitational force experienced by an observer standing on Earth, a.k.a massive object, is exactly equal to the pseudo-force an observer  experiences while abiding the non-inertial frame of reference of the freely falling elevator cabin. In effect, ''...it is impossible to speak of the absolute acceleration of the system of reference, just as the usual theory of relativity forbids us to talk of the absolute velocity of a system;''\cite{albert}  which leads to the hypothesis at the heart of general theory of relativity, namely the geodesic hypothesis. It states that test particles are  ''freely-falling'' and move along geodesic trajectories, which statement is solely derivable from the equivalence principle.   

Let us assume that there are no electro-magnetic fields present in space-time. Therefore, the first two force terms in (\ref{F}) vanish and (\ref{F_revealed}) becomes
\begin{eqnarray}\label{F_geodesic}
f^{j}=m \ddot{x}^{j} = m \Gamma^{j}_{\mu \nu}\dot{x}^{\mu} \dot{x}^{\nu}
\end{eqnarray}
the definition of a ''geodesic force'' an abserver standing on a massive object (which curves the geometry of space-time encoded by the metric $g_{\mu \nu}$) should include in the description of the motion of a test particle falling with respect to him.  In the absence of electro-magnetic fields the equations (\ref{eq.f1}) and (\ref{eq.dt g})  become
\begin{eqnarray}\label{F_geodesic}
 F^{\mu \nu} &=& - \frac{\hbar^2}{2m} \left(  \Gamma^{\nu}_{\; jk} \partial^{j}  g^{\mu k} - \Gamma^{\mu}_{\; jk} \partial^{j}  g^{\nu k} \right) \\
 i \hbar \partial_{t} g^{\mu \nu} &=& \frac{\hbar^2}{2m} \left(  \Gamma^{\nu}_{\; jk} \partial^{j}  g^{\mu k} + \Gamma^{\mu}_{\; jk} \partial^{j}  g^{\nu k} \right),
\end{eqnarray}

while (\ref{[x,x]}), (\ref{[x, dot x]}) and (\ref{form [dot x, dot x]}) remain
\begin{eqnarray}
\left[  x^{\mu} ,  x^{\nu}  \right]=0, \qquad
m \left[  x^{\mu} ,  \dot{x}^{\nu}  \right]=i \hbar g^{\mu \nu}
\end{eqnarray}
and
\begin{eqnarray} 
m  \left[  \dot{x}^{\mu}  ,  \dot{x}^{\nu}  \right] &=& 
F^{\mu \nu} + i \hbar \left(  \partial^{\nu}  g^{\rho \mu}  - \partial^{\mu} g^{\rho \nu}  \right) \dot{x}_{\rho}.
\end{eqnarray}
The listed commutational relations hold in curved space-time, that is in the vicinity of a massive object.

Note, where space-time is flat, the quantum mechanical commutational relations are given by the standard expressions
\begin{eqnarray}\label{commutator_flat}
\left[  x^{\mu} ,  x^{\nu}  \right]=0, \quad, 
m \left[  x^{\mu} ,  \dot{x}^{\nu}  \right]=i \hbar \delta^{\mu \nu} \quad {\rm and} \quad  \left[  \dot{x}^{\mu}  ,  \dot{x}^{\nu}  \right] &=& 0.
\end{eqnarray}
Therefore, the difference between the ''freely-falling'' quantum mechanical system (\ref{commutator_flat}) and the one forced to abide in the whareabouts of a massive object, that is the one experiencing the curved geometry of space-time is the r.h.s. of the cannonical commutational relations. Gravitation preserves localizability, but alters the operator algebra. In effect, the Heisenberg uncertainty relations between position and momentum operators (since they are proportional to the commutator) change:
$\Delta A \Delta B \geq \left| \langle \left[  \hat{A} , \hat{B}  \right] \rangle \right|/2.
$ The consequences for the precisions (the $\Delta$'s) of the simultaneous measurements of the position and momentum components are geometry dependent:
\begin{eqnarray}
\Delta  x^{\mu}  \Delta p^{\nu}  \geq \frac12 \hbar \left| g^{\mu \nu} \right|.
\end{eqnarray}

The alteration gravitation induces in the uncertainty relations and the operator algebra leads to yet another result: the change in the phase space volume occupied by an energy eigenstate. Indeed, the dimensionless volume measure $\delta \Gamma$ of the phase space
\begin{eqnarray}
d \Gamma =\prod_{i=1}^{N}  \frac{d x^{i} d p^{i}}{h}
\end{eqnarray}
increases in the presence of a gravitational field:
\begin{eqnarray}
d \Gamma \geq |g|,
\end{eqnarray}
where $g$ is the determinant of the metric. Since $\int d \Gamma$ is a measure of the number of energy eigenstates in the corresponding phase space volume, we can conclude that gravitation has the potential to change the number of quantum states in the part of phase space under consideration. As a result the phase space of the quantum system enlarges due to gravity ($|g| \geq 1$). Note, where space-time produces a singularity, that is $|g| \to \infty,$ the quantum system would have infinitely many energy eigenstates:
\begin{eqnarray}
\lim_{|g| \to \infty} d \Gamma = \infty
\end{eqnarray}
in the usual unit volume of phase space. This is an argument in support of the formation of singularities during the gravitational collapse of ordinary matter.

\section{The physical meaning of $\left[ {p}^{\mu}  ,   {p}^{\nu}  \right]\neq 0$}\label{sec.commutator}

An analogy with the commutational relations between the kinetic mementum components comes into mind (provided we assume the standard canonical commutational relations $[x,p]=i \hbar$ and $[p^\mu, p^\nu]=0$): $ 
 \left[ {p}^{\mu} - q A^{\mu}   ,   {p}^{\nu} - q A^{\nu}  \right]= i \hbar q \left( \partial^{\mu}  A^{\nu}  - \partial^{\nu}  A^{\mu}\right)= i \hbar q f^{\mu \nu},$ which is proportional to the Field strength tensor. Note, in the laboratory frame, it is the kinetic momentum that is actually measured

Next, one can use the field stength tensor formulation of the Maxwell's equation $
\partial_\mu f^{\mu \nu}= \mu_0 J^{\nu }\quad  J^{\nu }=(c\rho, \vec{j})$
and the continuity equation $\partial_\nu J^{\nu }= 0$ to define the field sources using the commutational relations:
\begin{eqnarray}\label{}
 {j}^{\;\nu } &=&  \frac{1}{i\hbar \mu_0 q} \partial_\mu \left[ {p}^{\mu}_{kin}    ,   {p}^{\nu}_{kin}  \right] 
\end{eqnarray}

The result is reinforced by returning to Dyson's report of Feynman's original attempt (1948) at deriving Maxwell equations\cite{dyson}. The approach yielded only the pair of equations which doesn't contain the sources.  Dyson's remark that ``the other two Maxwell equations (7) and (8) merely define the external charge and current densities'' lead to comments against the triviality of the statement\cite{comments}. 

In Dyson's paper one can notice something curious, that is upon combining equations (9) and (13) from \cite{dyson} leads to:
\begin{eqnarray}\label{*}
m^2 [\dot{x}_j, \dot{x}_k]&=&i \hbar \varepsilon_{jkl} H_l,
\end{eqnarray} 
where $H_l$ is the magnetic field. Now, take a partial derivative $\partial_k$ to get
\begin{eqnarray}\label{**}
\frac{1}{i \hbar}\partial_k [p_r, p_k]&=& \varepsilon_{rkl} \partial_k H_l={\rm rot}\vec{H}=j_r.
\end{eqnarray} 
Even more intriguing is the observation that (\ref{**}) contains the definition of charge density
\begin{eqnarray}\label{***}
\frac{1}{i \hbar} \left[ x_l , \partial_k [p_r, p_k] \right]&=& [x_l,j_r]=\rho [x_l, \dot{x}_j]=\rho \frac{i \hbar}{m} \delta_{rl}\\
\rho &=&-\frac{m}{\hbar^2} \left[ x_j , \partial_k [p_j, p_k] \right],
\end{eqnarray} 
which can be extracted when making the standard assumption for the current density $ j_r=\rho \dot{x}_r.$ A consistency relation based on the continuity equation/conservation of charge $-\partial_t \rho = {\rm div}\vec{j}$ emerges:
\begin{eqnarray}\label{****}
\frac{m}{\hbar^2} \partial_t \left[ x_j , \partial_k [p_j, p_k] \right]&=&\frac{1}{i \hbar}\partial_j \partial_k [p_j, p_k].
\end{eqnarray} 
However, ${\rm div } \vec{j}$ is vanishing due to symmetry considerations based on contracting the symmetric on continuous functions $\partial_\nu \partial_\mu$ operator and the anti-symmetric quantities in the momentum components' commutator $\left[  p^{\mu}  ,  p^{\nu}  \right] = 
m F^{\mu \nu} + G^{\mu \nu \rho} p_{\rho} :$
\begin{eqnarray}\label{}
{\rm div } \vec{j}&=&\partial_\nu j^{\nu } =  \frac{1}{i\hbar} \partial_\nu \partial_\mu \left( m F^{\mu \nu} + G^{\mu \nu \rho} p_{\rho}  \right) = 0. 
\end{eqnarray} 
Therefore $-\partial_t \rho = 0,$ that is $\rho$ is constant: 
\begin{eqnarray}\label{}
\rho &=&\frac{m}{(i\hbar)^2} \left[ x_j , \partial_k [p_j, p_k] \right] =\frac{m}{(i\hbar)^2} \partial_\nu G^{\mu \nu \rho} \left[ x_\mu ,  p_{\rho} \right]
+ \frac{m}{(i\hbar)^2} G^{\mu \nu \rho} \left[ x_\mu ,  \partial_\nu p_{\rho} \right] \\
&=&\left\{
\begin{array}{ll}
\frac{m}{i\hbar} \partial_\nu G^{\mu \nu \mu} & {\rm for} \quad \partial_\nu p_{\rho} = f_{\nu \rho}(\vec{r},t) \\
\frac{m}{i\hbar} \partial_\nu G^{\mu \nu \mu}  + \frac{m}{i\hbar}  G^{\mu \nu \rho}  W_{\nu \rho \mu}   & {\rm for} \quad  \partial_\nu p_{\rho} = f_{\nu \rho}(\vec{r},t) + W_{\nu \rho \tau}p^{\tau}   
\end{array}
\right.
\end{eqnarray} 
 which can be interpreted as the source of the electric field (charge) being the geometrical disterbance. Note, the mass $m$ of the quantum particle (system) multiplies the geometric term and renders mass-less charges non-existent. 

The result is not at all surprising and falls into the category of matter being described as the result of the curvature of space-time, i.e. geometrodynamics. The forefathers of this interpretation are Riemann and Clifford\cite{cliff}. Later Wheeler showed that a properly constructed ball of gravitational and electromagnetic radiation possesses the properties of a material body (called ''geon'')\cite{wheeler}. Most clearly and rigorously the  Campbell-Magaard theorem states that any n-dimensional Riemannian manifold (including matter) can be locally embedded in an (n + 1)-dimensional Ricci-flat Riemannian manifold, that is matter in 4D space-time can be thought of as pure geometry in 5D space-time\cite{campbell}.

\section{Quantum matter interaction with the curved geometry}\label{sec.G-QM int}

In mechanics, the volume of the phase space cells does not affect physical
quantities (heat capacity, pressure, ect.), provided the entropy of the system is proportional to the logarithm of the number of micro-states that correspond to the same macro-state. The chemical potential 
\begin{eqnarray}
\mu = - T \frac{\partial S}{\partial N} _{E, V = {\rm const.}},
\end{eqnarray}
on the other hand, depends of the phase space volume occupied by a micro-state and as a result the diffusion of particles along the gradient of densities or the chemical potential is a function of the elementary volume of the the phase space. 

In this paper, we explore the line of thought according to which the phase space volume can be deformed by gravitation and as a result the non-extensive
expression of the number of states should deviate from the one of a non-interacting system. Given the dependence of the chemical potential on the phase space cell volume, it is likely that gravitation interacts with the quantum system by changing the chemical potential or number of (quasi-) particles $N,$ which is a fundamental parameter conjugate to the chemical potential. In effect, gravitation changes the size of the quantum system and if one looks for a mechanism similar to the way matter affects the geometry of space-time, that is General Theory of Relativity on a quantum level, this mechanism should be the inverse of the just described. Quite likely, quantum matter can affect the geometry of space-time by changing its size (number of particles) which may involve a re-formulation of the particle number operator, its conjugate - the phase and their commutational relation.

The observation in mechanics that the size of a volume element in phase space remain constant in time (evolution parameter) is expressed as Liouville's theorem. However, this theorem no longer holds when there are momentum dependent forces such as the magnetic and geodesic one in (\ref{F}). In this case the phase space density is compressible.

Let us define a distribution function $f=f({\bf q, p},t)$ for a particular system. The fraction of systems $\delta N$ from an $N$ number of identical systems ensamble, which at time $t$ have coordinates
and momenta within an infinitesimal portion of the canonical phase space $[{\bf q}, {\bf q} + \delta {\bf q}] $ and $[{\bf p}, {\bf p}+ \delta {\bf p}]$ is used to define the phase space distribution
function  $f=f({\bf q, p},t)$  according to
\begin{eqnarray}
 \delta N &=& f({\bf q, p},t) \delta {\bf q} \delta {\bf p}.
\end{eqnarray}
Now, consider a small volume element of phase space, then the number of trajectories
entering a rectangular volume element through any face will in general
be different from the number which leave through an opposite face. 
Therefore, the equation of motion for the canonical phase space ${\bf \Gamma}=({\bf q, p})=(q_1, q_2, \cdots, q_{3N}, p_1, p_2, \cdots, p_{3N} )$ is given by an interpretation of the continuity equation and for the time rate of change of the distribution function one obtains
\begin{eqnarray}
\frac{d \ln f}{dt}= - \Lambda \left({\bf \Gamma}\right),
\end{eqnarray}
 where $\Lambda \left({\bf \Gamma}\right)= \frac{\partial }{\partial {\bf \Gamma}}\cdot \dot{{\bf \Gamma}}$ is the phase space compression factor\cite{EM}. The result is obtained without the use of the equations of motion, therefore its correctness is not related to the existence of a Hamiltonian. Provided the equations of motion can be derived from a Hamiltonian, then $\Lambda \left({\bf \Gamma}\right)=0$, identically regardless the presence of external fields which drive the system away from equilibrium.
Note, the existence of a Hamiltonian is a sufficient, but not necessary condition for $\Lambda \left({\bf \Gamma}\right)=0,$ that is for incompressible phase space (Liouville's equation). Interestingly,  the Liouville's theorem can be violated when
\begin{eqnarray}
\Lambda \left({\bf \Gamma}\right) \neq 0,
\end{eqnarray}
 which happens when:
 \begin{itemize}
 \item[--] sources or sinks of (quasi-) particles are present. Quantum matter may be interpreted as any quantum quasi-particle system. The most obvious examples being the super-fluid or superconducting condensate.
\item[--] existence of momentum dependent forces such that $ \nabla_p \cdot {\bf F} \neq 0 ,$ that is megnetic or geodesic ones in (\ref{F}). Therefore, it appears that gravitational interaction with a quantum system would necessarily require the compressibility of the phase space of that system, therefore gravitation can be viewed as more or less an information (entropic) signal due to its ability to changes probabalistic quantities.
\item[--] boundaries which lead to (quasi-) particle trapping or exclusion, which renders invalid the one-to-one mapping of distributions from one point to another. One possible interpretation of such a boundary is the insulating layer between the superconducting layers of a Josephson junction.
\item[--] spatial inhomogeneities that lead to velocity filtering, temporal variability at source or any process which leads to non-simultaneous observation of oppositely-directed trajectories.
\end{itemize}
In effect, gravitation introduces compressibility to the phase space volume $
{d \ln f}/{dt} \neq 0,$ which is its main coupling to the quantum system. Therefore, the inverse should also hold: {\it any quantum mechanical system or process which violates the incompressibility of the phase space couples to the gravitational field}. This constitutes the quantum matter-gravitation interaction within this framework.

\section{Conclusions}

We recognize Feynman's argument (in deriving the homogeneous Maxwell equations and the Lorentz force law from Newton's law of motion and the commutation relations between position
and momentum for a single non-relativistic particle) as an opportunity to use the quantum system as a test particle probing the dynamics of the underlying force-fields, gravitational in particular. It is recognized that gravitation affects quantum dynamics by modifying operator algebra. The emerging commutator between momentum's components is recognized as being proportional to electromagnetic field strength tensor. We define electromagnetic field sources through momentum's components commutator which is proportional to geometric (gravitational) quantities. We conclude that quantum matter - gravitation interaction takes place mainly in the phase space and any process which violates the compressibility of the quantum mechanical system's phase space can be recognized as gravitational interaction or induce gravitational response.

\section*{Acknowledgements}

The author declares that there are no confilict of interests regarding this paper and acknowledges partial financial support by the National Science Fund of Bulgaria under grant DN18/9-11.12.2017.

\end{document}